\documentclass[conference]{IEEEtran}
\IEEEoverridecommandlockouts

\usepackage{cite}
\usepackage{amsmath,amssymb,amsfonts}
\usepackage{algorithmic}
\usepackage{graphicx}
\usepackage{textcomp}
\usepackage{xcolor}
\usepackage{enumitem}
\usepackage{multirow}
\usepackage{makecell}
\usepackage{booktabs}

\usepackage[utf8]{inputenc}
\usepackage{url}
\def\BibTeX{{\rm B\kern-.05em{\sc i\kern-.025em b}\kern-.08em
    T\kern-.1667em\lower.7ex\hbox{E}\kern-.125emX}}
\begin{document}

\title{Polarity-Asymmetric Structural Calibration \\ for Link Sign Prediction}

\author{
  \IEEEauthorblockN{Qiqi Gao}
  \IEEEauthorblockA{Northeast Normal University \\Changchun, Jilin, China \\
  gaoqiqi777@nenu.edu.cn}
  \and
  \IEEEauthorblockN{Wenzhuo Song\textsuperscript{*}}
  \IEEEauthorblockA{Northeast Normal University
  \\Changchun, Jilin, China \\
  wzsong@nenu.edu.cn}
  \and
  \IEEEauthorblockN{Xueyan Liu}
  \IEEEauthorblockA{Jilin University
  \\Changchun, Jilin, China \\
  xueyanliu@jlu.edu.cn}
  \thanks{\textsuperscript{*}Corresponding author: Wenzhuo Song.}
}
\maketitle

\begin{abstract}
Link sign prediction (LSP) aims to infer the positive or negative polarity of unobserved links in signed networks. Signed Graph Neural Networks (SGNNs) usually rely on signed-graph structural priors, including structural balance and homophily-like similarity, to guide message passing and prediction. These priors describe population-level tendencies, not guarantees for individual target edges. Their failures are especially costly under severe sign imbalance, where errors on minority and locally conflicting relations are harder to detect and correct. We propose Polarity-Asymmetric Structural Calibration (PASC), a target-edge structural-prior calibration framework for signed link prediction. PASC constructs a structure-only prior representation, estimates a target-edge structural prior score, and compares this score with a local signed-context cue to derive a conflict residual. The residual calibrates signed attention aggregation, target-edge gated fusion, and regime-adaptive optimization. Experiments on five real-world signed network datasets show that PASC consistently achieves the best Macro-F1 among representative baselines, with competitive AUC, Binary-F1, and Micro-F1. Structural-shift experiments further suggest reduced dependence on dense-neighborhood and local-closure shortcuts. Source code is available at \url{https://github.com/iqqGGGGGGG/PASC-for-LSP}.

\end{abstract}


\section{Introduction}

Interactions between entities in real-world social networks often possess opposing polarities, such as trust/distrust and support/opposition \cite{Leskovec2010Sentiment,snapnets}. Such relations are modeled as signed networks, where edges encode positive or negative relational polarity rather than merely the existence of links. Link sign prediction (LSP) predicts the latent polarity of relations between node pairs, with applications in recommender systems \cite{cao2026recommendation,wang2026signed}, sentiment analysis \cite{zhao2024Sentiment}, and community detection \cite{he2022sssnet}. Signed Graph Neural Networks (SGNNs) can capture complex signed topological patterns and have become the dominant approach to LSP \cite{song2018learning,derr2018signed,huang2021sdgnn,zhang2023rsgnn}.

\begin{figure}[htbp]
  \centering
  \includegraphics[width=1\linewidth]{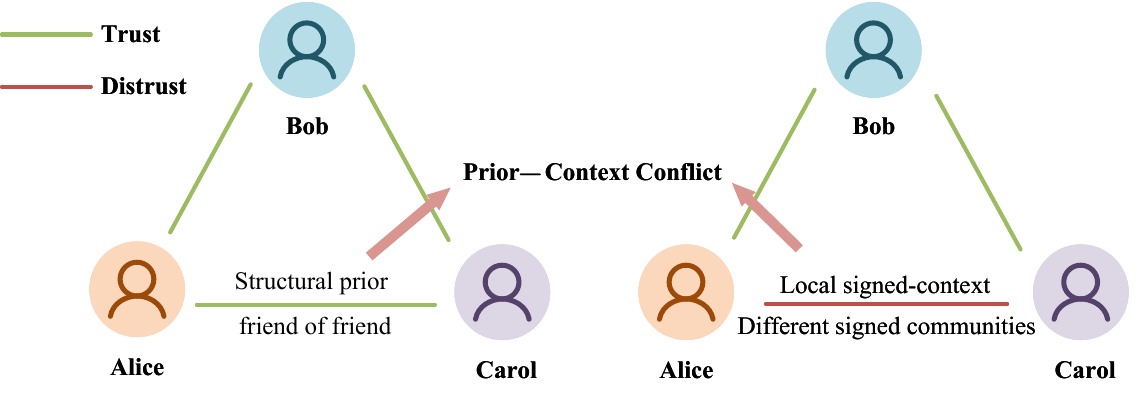}
  \caption{Example of target-edge structural-prior unreliability. Alice and Carol share a positive neighbor Bob, so structural balance suggests a positive relation. Their local signed context, however, may provide conflicting evidence. Structural priors capture population-level tendencies but need not be reliable for every target edge.}
  \label{fig:1}
\end{figure}

Existing SGNNs often use recurring signed graph patterns as structural priors. Two examples are structural balance \cite{Heider1946Theory,Dorwin1956Theory}, where triangle signs provide cues for the target edge, and homophily \cite{Miller2001Homophily,Gueorgi2009Homophily}, where similar nodes are more likely to form positive links. These priors are usually implemented through signed message passing, separate aggregation over positive and negative neighbors, or higher-order path and subgraph features \cite{derr2018signed,huang2021sdgnn,xu2023spmf,fang2026collaborative}. Such designs encode useful topological evidence beyond individual node features.

However, structural priors describe recurring patterns across many edges rather than guarantees for every target edge.
Figure~\ref{fig:1} illustrates a case where a structural prior and local signed-context evidence point in different directions for the same target edge. Table~\ref{tab:prior_posterior} further shows that structural balance predictions do not always match actual edge polarities, following the analysis of Kim et al.~\cite{kim2025trustworthiness}. When the prior signs are $(+, -)$, balance theory predicts a negative posterior sign, yet 70.8\% of the corresponding edges in Wiki-RfA are actually positive. The proportions are 63.5\% and 53.5\% on Bitcoin-Alpha and Epinions, respectively. Thus, the same structural pattern can map to different polarities. Structural priors remain useful, but their reliability varies across target edges.

\begin{table*}[t]
\centering
\caption{Prior-Posterior Statistics of Triangles in Signed Networks}
\label{tab:prior_posterior}
\setlength{\tabcolsep}{7pt}
\renewcommand{\arraystretch}{1.3}
\begin{tabular}{cc|cc|cc|cc}
\hline
\multirow{2}{*}{\textbf{Triads}} & \textbf{Prior} & \multicolumn{2}{c|}{$(+, +)$} & \multicolumn{2}{c|}{$(+, -)/(-, +)$} & \multicolumn{2}{c}{$(-, -)$} \\
 & \textbf{Posterior} & $+$ & $-$ & $+$ & $-$ & $+$ & $-$ \\
\hline
\multirow{5}{*}{Datasets} & Bitcoin-Alpha & 95.7\% & 4.3\% & 63.5\% & 36.5\% & 86.6\% & 13.4\% \\
 & Bitcoin-OTC & 95.9\% & 4.1\% & 39.3\% & 60.7\% & 90.2\% & 9.8\% \\
 & Wiki-RfA & 88.1\% & 11.9\% & 70.8\% & 29.2\% & 66.9\% & 33.1\% \\
 & Slashdot & 94.9\% & 5.1\% & 46.5\% & 53.5\% & 68.5\% & 31.5\% \\
 & Epinions & 96.5\% & 3.5\% & 53.5\% & 46.5\% & 69.7\% & 30.3\% \\
\hline
\multicolumn{2}{c|}{\textbf{Balanced?}} & $\bigcirc$ & $\times$ & $\times$ & $\bigcirc$ & $\bigcirc$ & $\times$ \\
\hline
\end{tabular}
\end{table*}

Polarity imbalance amplifies the impact of prior-related errors during learning and evaluation. As shown in Table~\ref{tab:datasets}, positive edges are 3.4 to 14.7 times more frequent than negative edges across the five datasets. Under this imbalance, a model can fit structurally typical and majority-polarity relations while errors on minority negative or locally conflicting target edges contribute less to the training objective and are less visible in global metrics such as AUC or Micro-F1. The key challenge is therefore to estimate when a structural prior should be trusted for a specific target edge and to calibrate its influence under the local signed context.

Most existing methods either extract richer structural evidence through multi-hop propagation, motif modeling, or subgraph encoding, or reduce unreliable structures through denoising, filtering, or reweighting \cite{xu2023spmf,fang2023signed,fang2026collaborative,zhang2023rsgnn,kim2025trustworthiness,li2025Self}. These strategies improve signed graph representation learning, but they mainly treat structural reliability as a property of patterns or edges to be extracted, suppressed, or reweighted. They do not directly decide how strongly each target edge should rely on a structural prior under its own local signed context.

We propose Polarity-Asymmetric Structural Calibration (PASC) for link sign prediction. PASC treats signed-graph patterns as useful but imperfect priors and calibrates their reliability for each target edge under its local signed context. It first uses truncated singular value decomposition to construct a structure-only prior representation, from which target-edge structural prior scores are estimated. It then derives local signed-context cues from signed clustering \cite{he2022sssnet} and computes the conflict residual between the prior score and the local signed-context cue. This residual guides signed attention aggregation \cite{Huang2019sigat} and calibrates neighbor message propagation according to the prior-context mismatch. PASC then fuses base and context representations through a target-edge gate, allowing each target edge to balance structural and contextual information.

The same conflict residual also shapes the training objective through two variants suited to different graph scales. In small-scale graphs, where local signed-context evidence is limited, the soft variant PASC-S reweights training edges to emphasize larger prior-context mismatch. In large-scale graphs, where enough candidate edges are available for stronger supervision, the harder variant PASC-H pairs structurally matched but context-opposite edges as residual-contrastive supervision. On five real-world signed network datasets, PASC consistently achieves the best Macro-F1 while maintaining competitive AUC, Binary-F1, and Micro-F1. Under Degree-Shift and Structural-Shortcut-Shift, PASC also shows reduced dependence on structural shortcuts.

Our contributions are summarized as follows:
\begin{itemize}
    \item Rather than encoding richer structure, we introduce a target-edge structural-prior calibration perspective for imbalanced signed link prediction, asking when a model should trust structural priors for each target edge and calibrating their influence under the local signed context.
    \item We define the conflict residual, an observed-minus-expected measure between a structural-gradient-based prior score and a local signed-context cue, as the core signal that drives calibration. The structural gradient provides a lightweight, training-free structural signal for this calibration.
    \item Driven by the same conflict residual, we instantiate the calibration consistently at three levels, namely residual-guided signed attention for aggregation, target-edge gated fusion for representation fusion, and regime-adaptive conflict-residual optimization for the training objective, realized as residual-aware reweighting (PASC-S) and residual-contrastive supervision (PASC-H).
    \item Experiments on five benchmarks show the best Macro-F1 and competitive overall performance, with additional structural-shift evidence.
\end{itemize}

\section{Related Work}

We review three lines of related work: signed graph neural networks, higher-order and subgraph-based methods, and graph debiasing techniques.

\subsection{Signed Graph Neural Networks}\label{2.1}

Signed graph representation learning often incorporates signed-graph structural priors into representation objectives or message-passing mechanisms. Early methods such as SiNE \cite{Wang2017SignedNE} and nSNE \cite{song2018learning} encode social status and structural balance theories through distance-based optimization objectives. Beyond Euclidean embedding methods, HSNE embeds signed networks
in a Poincaré ball and combines structural balance theory with
Riemannian optimization to capture latent hierarchical structure
\cite{SONG2021329}. SGCN \cite{derr2018signed} introduces dual-channel aggregation to separately model positive and negative neighbors. Subsequent models, including SiGAT \cite{Huang2019sigat} and SDGNN \cite{huang2021sdgnn}, further exploit attention mechanisms and directed closed-loop structures to capture richer signed relational patterns.

These architectures are effective, but their aggregation mechanisms can still overemphasize dominant structural patterns under polarity asymmetry. Minority or structurally conflicting edges are then harder to model \cite{kim2025trustworthiness}, even when global ranking metrics remain high \cite{liu2023imbalanced, imani2026roc}.

\subsection{Subgraph and High-Order Heuristic Methods}\label{2.2}
Another line of work enhances signed graph representations with higher-order or subgraph-level patterns, including multi-order adjacency matrices, motifs, and closed subgraphs \cite{xu2023spmf, fang2023signed, fang2026collaborative, Chen2024Motif}. SPMF \cite{xu2023spmf} quantifies multi-hop paths from structural equilibrium theory and uses multi-order signed adjacency matrices. SELO \cite{fang2023signed} extracts closed subgraphs through local breadth-first search and characterizes their balanced or unbalanced structures. CFSE \cite{fang2026collaborative} integrates global collaborative filtering signals with local subgraph features. Although these methods improve structural expressiveness, explicit higher-order or subgraph encoding adds computational overhead and scalability challenges \cite{Zhang2025Power}, especially when target-pair subgraphs must be repeatedly constructed. Minority-polarity or locally heterogeneous regions may also provide too few higher-order patterns.

\subsection{Imbalanced Learning and Debiasing on Graphs}\label{2.3}
As signed graph representation learning moves to noisy or distribution-shifted settings, robustness has become a recurring concern \cite{Chao2026causal,Liu2021Tail,Jin2021Defenses}. In signed graphs, several methods reduce the influence of unreliable structures. RSGNN \cite{zhang2023rsgnn} filters edges that violate topological rules using a structural regularizer grounded in structural balance theory. SE-SGformer \cite{li2025Self} identifies positive and negative neighbors most relevant to the target node. TrustSGCN \cite{kim2025trustworthiness} assigns trust scores from local topological features and global statistical ratios, restricting propagation through low-scoring edges.

These methods improve robustness by filtering, reweighting, or explaining structural information. In imbalanced signed networks, however, edges that deviate from common structural priors may be informative minority or locally conflicting relations rather than noise. Treating them mainly as unreliable structures can weaken class-balanced prediction, which motivates our target-edge calibration perspective.

\section{Methodology}
\begin{figure*}[htbp]
  \centering
  \includegraphics[width=1\linewidth]{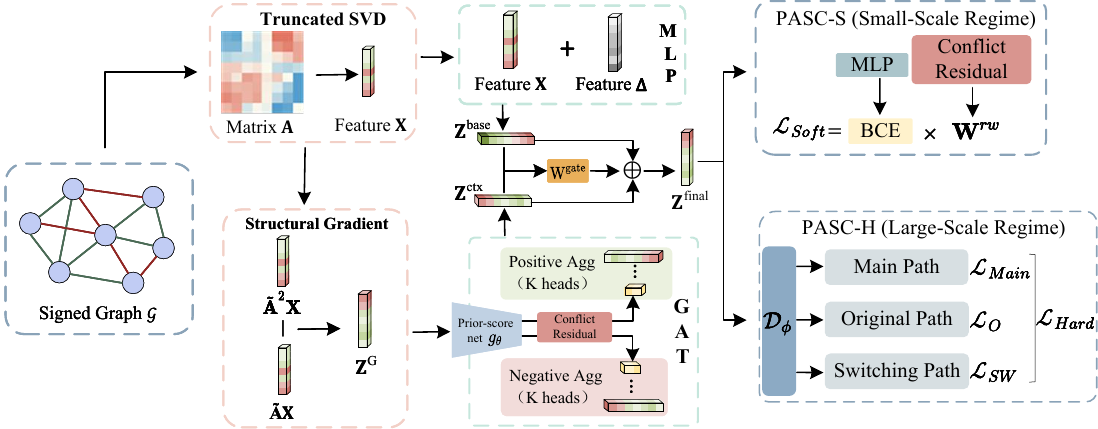}
    \caption{Architecture of the PASC framework for link sign prediction. PASC estimates target-edge structural prior scores, derives conflict residuals from local signed-context cues, and uses residual-guided aggregation, regime-adaptive conflict-residual optimization (PASC-S and PASC-H), and target-edge gated fusion to calibrate structural reliance under different graph regimes.}
  \label{fig:2}
\end{figure*}

In this section, we present the PASC framework, as illustrated in Fig.~\ref{fig:2}. PASC calibrates target-edge reliance on signed-graph structural priors by estimating a target-edge structural prior score, comparing it with a local signed-context cue, and using the resulting conflict residual to guide message aggregation and feature fusion. The same conflict residual drives all components of PASC, applying one consistent calibration signal at the aggregation, fusion, and optimization levels. Throughout, PASC estimates its calibration signals at the edge level but injects them into node-level message passing, and the resulting node representations are then reassembled into target-edge representations for prediction.

\subsection{Problem Definition}\label{3.1}

A signed graph is defined as $\mathcal{G} = (\mathcal{V}, \mathcal{E}^+, \mathcal{E}^-)$, where $\mathcal{V} = \{v_1, \ldots, v_N\}$ is a collection of $N$ nodes, and $\mathcal{E}^+$ and $\mathcal{E}^-$ are the sets of positive and negative links, respectively. For ease of analysis, we use an adjacency matrix $\mathbf{A} \in \mathbb{R}^{N \times N}$ to represent a signed graph, where $A_{ij} = 1$ indicates a positive link, $A_{ij} = -1$ indicates a negative link, and $A_{ij} = 0$ indicates no observed link between nodes $v_i$ and $v_j$. Given a signed graph $\mathcal{G} = (\mathcal{V}, \mathcal{E}^+, \mathcal{E}^-)$, the LSP task is to predict the link sign between a target node pair $(u,v)$. For datasets originally provided as directed edges, the current implementation uses an undirected signed-graph representation: each observed signed edge is symmetrized before the adjacency matrix is constructed, and edge direction is not modeled.

\subsection{Structural Prior Estimation and Residual-Guided Aggregation}\label{3.2}

This part of PASC builds the conflict residual and uses it to calibrate message passing. We first estimate a structural prior score for each edge from structure-only features, which captures the population-level structural tendency of an edge. We then define the conflict residual as the gap between this prior score and the observed local signed context, and use it as an edge-level bias that guides positive and negative signed attention aggregation, so that message passing is adjusted according to where the structural prior is unreliable.

\subsubsection{Structure-Only Prior Representation}\label{3.2.1}

To estimate the structural prior score of each target edge, we first construct structure-only node features $\mathbf{X}$ from the signed adjacency matrix, which serve as the substrate for the structural-gradient computation below. Here, structure-only means free of target-edge labels and local context, not free of edge signs.

We approximate the signed adjacency matrix using truncated singular value decomposition:
\begin{equation}
    \mathbf{A} \approx \mathbf{U}_r \mathbf{\Sigma}_r \mathbf{V}_r^\top,
\end{equation}
where $r$ is the rank, $\mathbf{\Sigma}_r \in \mathbb{R}^{r \times r}$ is the diagonal matrix of the top singular values, and $\mathbf{U}_r, \mathbf{V}_r \in \mathbb{R}^{N \times r}$ are the corresponding left and right singular matrices. We retain the top-$r$ structural bases as the initial node feature matrix:
\begin{equation}
    \mathbf{X} = \mathbf{U}_r \mathbf{\Sigma}_r^{1/2}.
\end{equation}

In practice, we exploit the sparsity of signed networks and compute the decomposition using Lanczos-based sparse solvers or randomized low-rank SVD. The resulting complexity scales approximately linearly with the number of observed edges, making the preprocessing practical for large graphs.

The structure-only features $\mathbf{X}$ summarize each node's position in the dominant structure of the signed graph, but they do not directly express how sharply the local neighborhood changes around a node. To capture this local structural variation, we compare the structure-only features aggregated from one-hop and two-hop neighborhoods, and define the structural gradient of each node as:
\begin{equation}
    \begin{aligned}
    \mathbf{Z}^{G} &= \tilde{\mathbf{A}}^2 \mathbf{X} - \tilde{\mathbf{A}} \mathbf{X} \\
    &= (\mathbf{D}^{-1} |\mathbf{A}|)^2 \mathbf{X} - (\mathbf{D}^{-1} |\mathbf{A}|) \mathbf{X},
    \end{aligned}
\end{equation}
where $\tilde{\mathbf{A}} = \mathbf{D}^{-1}|\mathbf{A}|$ denotes the row-normalized absolute adjacency matrix, $|\mathbf{A}|$ is the absolute signed adjacency matrix, and $\mathbf{D}$ is its degree matrix. We keep the signed adjacency for the SVD to preserve polarity-aware structural bases, while $\tilde{\mathbf{A}}$ uses $|\mathbf{A}|$ to measure neighborhood smoothness independent of edge polarity. The structural gradient $\mathbf{Z}^{G}$ describes local structural variation: $\mathbf{Z}^{G}\approx\mathbf{0}$ indicates a locally smooth and stable neighborhood, whereas a large $|\mathbf{Z}^{G}|$ indicates a structural transition region, such as a community boundary or a bridging node. 

In such transition regions, neighborhoods are more diverse and balance- and homophily-type priors become less reliable across target edges. The structural gradient therefore serves as a lightweight proxy for where structural-prior reliability is less predictable.To estimate the structural prior score of a target edge $(u,v)$, we concatenate their structural-gradient representations and feed the joint representation into a prior-score network:
\begin{equation}
    e_{uv} = \sigma(\text{MLP}_{prior}([\mathbf{z}_u^{G} \parallel \mathbf{z}_v^{G}])),
\end{equation}
where $\text{MLP}_{prior}$ is a multi-layer perceptron, and $e_{uv} \in [0,1]$ is a learnable target-edge structural prior score. It is computed only from structural features, without taking the observed local signed context or the target edge's own sign label as input. $\text{MLP}_{prior}$ is trained end-to-end with the rest of PASC under the link sign prediction objective, with no separate supervision from $T_{uv}^{f}$.

\subsubsection{Residual-Guided Attention Aggregation}\label{3.2.2}

Residual-guided attention realizes this calibration during message passing. Graph attention typically weights neighbors by feature similarity. In signed graphs, this favors typical, structurally consistent neighbors and underweights the minority or sign-conflicting edges that are often the most informative for sign prediction. To calibrate this process, PASC computes a conflict residual for each edge and uses it as an edge-level attention bias during node-level message passing. The conflict residual contrasts the structural prior score $e_{uv}$ estimated above with the edge's observed local signed context, so that a large gap signals that the structural prior is unreliable for that edge. To obtain this observed context, we apply a signed graph clustering algorithm \cite{he2022sssnet}, which assigns a community label $c(\cdot)$ to each node. For a target edge $(u,v)$, we define the observed context as:
\begin{equation}
    T_{uv}^f = \mathbb{1}[c(u) = c(v)],
\end{equation}
where $\mathbb{1}[\cdot]$ is the indicator function. Thus, $T_{uv}^f=1$ if $u$ and $v$ belong to the same community, and $T_{uv}^f=0$ otherwise. We then compute the conflict residual between the local signed-context cue and the structural prior score:
\begin{equation}
    R_{uv} = T_{uv}^f - e_{uv}.
\end{equation}

During message passing, we evaluate this residual on each observed aggregation edge $(i,j)$ in the training graph, where $j$ is a signed neighbor of the center node $i$. The conflict residual $R_{ij}$ then acts as an attention bias: a value close to zero indicates that the local context is consistent with the structural expectation, whereas a larger absolute value indicates stronger mismatch. We use it as a calibration signal for attention aggregation:
\begin{equation}
\begin{split}
\alpha_{ij}^{(k)} = \text{Softmax}_j \Big( &\mathbf{a}_k^T \text{LeakyReLU}\big( \mathbf{W}^{(k)} [\mathbf{z}^{base}_i \parallel \mathbf{z}^{base}_j] \big) \\
&+ \phi_{bias}(R_{ij})^{(k)} \Big),
\end{split}
\end{equation}
where $\mathbf{W}^{(k)}$ and $\mathbf{a}_k$ are learnable parameters for the $k$-th attention head, and the scalar residual is mapped to a head-specific additive bias $\phi_{bias}(R_{ij})=\tanh(\mathbf{W}_r R_{ij}+\mathbf{c}_r)\in\mathbb{R}^{K}$, whose $k$-th entry $\phi_{bias}(R_{ij})^{(k)}$ biases head $k$. This calibrates the attention score according to the mismatch between the structural prior score and the local signed context. The base node representations $\mathbf{z}^{base}_i$ and $\mathbf{z}^{base}_j$ are encoded from the structure-only features by adding a small learnable perturbation $\mathbf{\Delta}$ before an MLP:
\begin{equation}
    \mathbf{Z}^{base} = \text{MLP}_{base}(\mathbf{X} + \gamma \mathbf{\Delta}),
\end{equation}
where $\gamma$ is the scaling coefficient that controls the intensity of the learnable perturbation $\mathbf{\Delta}$.

For each node $i$, we apply residual-guided attention aggregation on its positive and negative neighborhoods separately. The two aggregated messages are concatenated and projected to obtain the context representation of node $i$:
\begin{equation}
\begin{split}
\mathbf{z}_i^{ctx} = \text{MLP}_{ctx} \Big( \Big[
&\sum_{j \in \mathcal{N}_i^+} \alpha_{ij}^{+} (\mathbf{W}_s \mathbf{z}_j^{base}) \\
\parallel\;&\sum_{j \in \mathcal{N}_i^-} \alpha_{ij}^{-} (\mathbf{W}_s \mathbf{z}_j^{base})
\Big] \Big),
\end{split}
\end{equation}
where $\mathcal{N}_i^+$ and $\mathcal{N}_i^-$ denote the sets of positive and negative neighbors of node $i$, respectively, $\alpha_{ij}^{+}$ and $\alpha_{ij}^{-}$ are the residual-calibrated multi-head attention weights over the positive and negative neighborhoods, and $\mathbf{W}_s$ is a shared feature projection matrix.

\subsection{Target-Edge Gated Fusion}\label{3.3}
The base representation $\mathbf{z}^{base}$ encodes the structural-prior view of a node, whereas the context representation $\mathbf{z}^{ctx}$ carries the residual-calibrated local signed context. How much a target edge should rely on its local context depends on whether its structural prior is reliable, which is exactly what the conflict residual measures. Rather than fixing this balance, PASC extends this calibration to the fusion level with a target-edge gate. Unlike the neighbor-normalized attention weights used above, the gate is a single scalar in $(0,1)$ that scales the residual-calibrated context $\mathbf{z}^{ctx}$ before it is added to the structural representation. A larger gate injects more of the calibrated local context, whereas a smaller gate keeps the fused representation closer to the structural prior. For a target edge $(u,v)$, the gate takes the base and context representations of both endpoints as input and computes the fusion weight:
\begin{equation}
    w_{uv}^{gate} = \sigma(\text{MLP}_{gate}([\mathbf{z}_u^{base} \parallel \mathbf{z}_v^{base} \parallel \mathbf{z}_u^{ctx} \parallel \mathbf{z}_v^{ctx}])).
\end{equation}

Although $w_{uv}^{gate}$ is computed for the target edge $(u,v)$, it controls how much context is injected into the base representations of its two endpoint nodes. We first obtain the node-level final representations as:
\begin{equation}
\begin{aligned}
\mathbf{z}_u^{final} &= \mathbf{z}_u^{base} + w_{uv}^{gate} \cdot \mathbf{z}_u^{ctx},\\
\mathbf{z}_v^{final} &= \mathbf{z}_v^{base} + w_{uv}^{gate} \cdot \mathbf{z}_v^{ctx}.
\end{aligned}
\end{equation}

The two node-level final representations are then concatenated to form the edge-level prediction input:
\begin{equation}
    \mathbf{z}_{uv}^{edge} = [\mathbf{z}_u^{final} \parallel \mathbf{z}_v^{final}].
\end{equation}

\subsection{Regime-Adaptive Conflict-Residual Optimization}\label{3.4}

The aggregation and fusion modules above use the conflict residual to calibrate message passing and target-edge representation fusion. At the optimization level, PASC uses the same residual to decide which edges should receive stronger supervision. Edges with a large $|R_{uv}|$ indicate a stronger mismatch between the structure-derived prior and the observed local signed context, and thus carry more informative signal for learning when structural priors should be trusted. We instantiate this idea through two strategies whose stability depends on the graph regime. In small-scale graphs, where local evidence is limited and strong calibration can amplify variance, PASC-S applies soft residual-aware reweighting. In large-scale graphs, where the abundance of edges provides enough structurally similar but context-opposite candidates, PASC-H adds residual-contrastive twin supervision. Graph scale therefore serves as a proxy for candidate availability, which determines which variant to use.

\subsubsection{Soft Reweighting for Small-Scale Signed-Topology Regimes}
For small-scale graphs, PASC-S performs sample reweighting with the structural prior score $e_{uv}$ while keeping the graph topology unchanged. To avoid unstable weights when $e_{uv}$ is close to 0 or 1, we first apply a truncation bound:
\begin{equation}
    \hat{e}_{uv} = \max(\epsilon, \min(e_{uv}, 1-\epsilon)),
\end{equation}
where $\epsilon$ denotes the truncation boundary. Given the local signed-context cue $T_{uv}^f$, we compute the edge-level reweighting coefficient as:
\begin{equation}
    w_{uv}^{rw} = \frac{T_{uv}^f}{\hat{e}_{uv}} + \frac{1-T_{uv}^f}{1-\hat{e}_{uv}}.
\end{equation}

The coefficient $w_{uv}^{rw}$ implements a soft prior-context mismatch reweighting strategy. The predictor maps the edge-level prediction input to the predicted soft label:
\begin{equation}
    \hat{y}_{uv}^{soft} = \text{MLP}(\mathbf{z}_{uv}^{edge}).
\end{equation}

The weighted observation loss is defined as:
\begin{equation}
    \mathcal{L}_{Soft} = \frac{1}{|\mathcal{E}_{train}|} \sum_{(u,v) \in \mathcal{E}_{train}} w_{uv}^{rw} \cdot \text{BCE}(\hat{y}_{uv}^{soft}, Y^{f}_{uv}),
\end{equation}
where $\mathcal{E}_{train}$ denotes the training edge set, $Y^{f}_{uv}$ denotes the observed training label of edge $(u,v)$, and $\text{BCE}(\cdot)$ is the binary cross-entropy loss. In terms of the conflict residual, $w_{uv}^{rw}$ grows monotonically with $|R_{uv}|$. When $T_{uv}^f=1$, a smaller $\hat{e}_{uv}$ gives a larger $|R_{uv}|$ and a larger weight, and when $T_{uv}^f=0$, a larger $\hat{e}_{uv}$ has the same effect. PASC-S can thus be interpreted as a soft residual-aware objective that emphasizes edges where the structural prior is less reliable. The reweighting does not treat $T_{uv}^f$ as a ground-truth sign. It only uses the disagreement between the structural prior score and the signed-community context as an indicator of structural unreliability.
\subsubsection{Residual-Contrastive Twin Supervision for Large-Scale Signed-Topology Regimes}

In large-scale regimes, reweighting alone may produce highly uneven weights and unstable optimization. PASC-H instead adds auxiliary supervision built from structural twins, using the structural prior score as a matching anchor to find edges with a similar prior score but the opposite local context. For a target edge $(u,v)$ with local signed-context cue $T_{uv}^{f}$, we construct the candidate set:
\begin{equation}
\mathcal{C}_{cand} = \{ (u', v') \mid T_{u'v'}^{f} = 1-T_{uv}^f \land |e_{uv} - e_{u'v'}| < \delta \},
\end{equation}
where $(u',v')$ is a candidate edge pair, $e_{uv}$ and $e_{u'v'}$ are the structural prior scores of the target edge and the candidate pair, respectively, and $\delta$ is the tolerance threshold for structural-prior matching. This candidate filtering restricts the search to pairs with similar structural prior score but opposite observed context. If no candidate satisfies the constraint, the target edge is used as its own fallback.

Within the candidate set, we select the structural twin that is most similar to the target edge in both initial structural features and structural-gradient representations. We define the edge-level feature concatenations as $\mathbf{X}_{uv} = [\mathbf{X}_u \parallel \mathbf{X}_v]$ and $\mathbf{Z}_{uv}^{G} = [\mathbf{Z}_u^{G} \parallel \mathbf{Z}_v^{G}]$, and compute the matched pair as:
\begin{equation}
(u', v') = \mathop{\arg\min}_{(u', v') \in \mathcal{C}_{cand}} 
\frac{1}{2} \Big( d(\mathbf{X}_{uv}, \mathbf{X}_{u'v'}) + d(\mathbf{Z}_{uv}^{G}, \mathbf{Z}_{u'v'}^{G}) \Big),
\end{equation}
where $d(\cdot,\cdot)$ denotes a distance function used to measure pairwise feature similarity.

After obtaining the matched structural twin $(u', v')$, we use its observed polarity as the switched-context label $Y_{u'v'}^f$. A conditional decoder $\mathcal{D}_\phi$ is then used to construct two auxiliary supervision paths that share the same endpoint base representations but differ in the context fed to the decoder and the supervised label.

In the original-context path, the decoder uses the edge's own context $T_{uv}^{f}$ to predict its observed label $Y^{f}_{uv}$:
\begin{equation}
\mathcal{L}_{O} = \text{BCE}(\mathcal{D}_\phi([\mathbf{z}^{base}_u \parallel \mathbf{z}^{base}_v \parallel T_{uv}^{f}]), Y^{f}_{uv}).
\end{equation}

In the switched-context path, the decoder is supervised by the twin label $Y_{u'v'}^f$ under the switched context $T^{f}_{u'v'}$:
\begin{equation}
\mathcal{L}_{SW} = \text{BCE}(\mathcal{D}_\phi([\mathbf{z}^{base}_u \parallel \mathbf{z}^{base}_v \parallel T^{f}_{u'v'}]), {Y}_{u'v'}^f).
\end{equation}

For the main prediction task, the edge-level prediction input is combined with the original context to predict the observed label:
\begin{equation}
\mathcal{L}_{main} = \text{BCE}(\mathcal{D}_\phi([\mathbf{z}_{uv}^{edge} \parallel T_{uv}^{f}]), Y^{f}_{uv}).
\end{equation}

The total loss of PASC-H is:
\begin{equation}
    \mathcal{L}_{Hard} = \mathcal{L}_{main} + \lambda_1 \mathcal{L}_{O} + \lambda_2 \mathcal{L}_{SW},
\end{equation}
where $\lambda_1$ and $\lambda_2$ control the strengths of the original-context and switched-context auxiliary losses. By pairing samples with similar structural prior scores but opposite local contexts, PASC-H constructs a residual-contrastive supervision signal in which the prior estimate is approximately held fixed while the observed context changes. This encourages the decoder to model how edge polarity varies with prior--context mismatch, rather than relying on the structural prior alone, improving robustness in large-scale signed graphs.

\textbf{Complexity.} All structure-dependent preprocessing is performed on the training graph. With sparse adjacency, the truncated SVD and the structural gradient both cost $O(r|\mathcal{E}_{train}|)$, where $r$ is the rank. Residual-guided signed attention keeps the asymptotic cost of standard signed attention, adding only a scalar bias per aggregation edge, while PASC-H bounds candidate matching with the threshold $\delta$ using score buckets.

\section{Experiments}
We evaluate PASC by answering the following five research questions:
\begin{itemize}
    \item RQ1: How does PASC perform in terms of overall performance and class-balanced prediction compared with representative signed link prediction methods?
    \item RQ2: How do the two residual-optimization variants behave across small-scale and large-scale signed-graph regimes?
    \item RQ3: How do residual-guided aggregation, target-edge gated fusion, and regime-adaptive optimization contribute to PASC?
    \item RQ4: How sensitive is PASC-H to the weights of original-context and switched-context auxiliary supervision?
    \item RQ5: How robust is PASC under structural-shift protocols?
\end{itemize}

\subsection{Experimental Setup}

\textbf{Datasets}. We use five real-world signed graph benchmarks. Bitcoin-Alpha and Bitcoin-OTC \cite{Kumar2016data} are user trust networks from Bitcoin trading platforms, where edges denote trust (positive) or distrust (negative) ratings. Wiki-RfA \cite{West2014data} is derived from the Wikipedia Request for Adminship process, where edges represent support (positive) or opposition (negative) votes. Slashdot and Epinions \cite{Leskovec2010Sentiment} are technology news and consumer review networks, respectively, where edges indicate explicit trust/friendship or distrust/enmity relations. As shown in Table~\ref{tab:datasets}, these datasets vary substantially in graph scale and exhibit clear polarity imbalance, with positive-to-negative edge ratios ranging from 3.4:1 to 14.7:1. These properties let us evaluate PASC across different scales and degrees of polarity imbalance.

\begin{table}[htbp]
  \caption{Statistics of the datasets used in our experiments.}
  \label{tab:datasets}
  \centering
  \small
  \setlength{\tabcolsep}{3pt}
  \begin{tabular}{lrrrrr}
    \toprule
    Dataset & Nodes & Edges & Pos links & Neg links & Pos/Neg \\
    \midrule
    Bitcoin-Alpha\textsuperscript{1} & 3,782 & 24,185 & 22,649 & 1,536 & 14.7:1 \\
    Bitcoin-OTC\textsuperscript{1}  & 5,881 & 35,591 & 32,028 & 3,563 & 9.0:1 \\
    Wiki-RfA\textsuperscript{1}     & 11,259 & 178,096 & 138,813 & 39,283 & 3.5:1 \\
    Slashdot\textsuperscript{1}    & 84,140 & 549,201 & 425,071 & 124,130 & 3.4:1 \\
    Epinions\textsuperscript{1}    & 131,827 & 841,371 & 717,667 & 123,704 & 5.8:1 \\
    \bottomrule
  \end{tabular}

  \makebox[\linewidth][l]{\footnotesize $^{1}$\,\url{https://snap.stanford.edu/data/}}
\end{table}

\textbf{Baselines}. We compare PASC with eight representative baselines (nSNE \cite{song2018learning}, SGCN \cite{derr2018signed}, SNEA \cite{li2020learning}, SDGNN \cite{huang2021sdgnn}, RSGNN \cite{zhang2023rsgnn}, SPMF \cite{xu2023spmf}, TrustSGCN \cite{kim2025trustworthiness}, and CFSE \cite{fang2026collaborative}) that cover embedding-based methods, signed message-passing models, multi-hop and subgraph-enhanced methods, denoising-based robust SGNNs, and trustworthiness-based propagation. For a fair comparison, we follow the reported settings of the original papers when available and tune key hyperparameters on the validation set using the same splits. SE-SGformer requires a substantially different transformer-style training pipeline and a self-explanation module that is not directly compatible with our unified SGNN evaluation protocol, so we leave its direct comparison to future extended evaluation.

\textbf{Evaluation Protocol}. All datasets are split into training, validation, and test sets in an 8:1:1 ratio. The current evaluation follows a static and transductive setting: the node set is fixed across the training, validation, and test edge splits. We use polarity-stratified edge splits so that the positive/negative ratio is approximately preserved across training, validation, and test sets. Following \cite{derr2018signed,huang2021sdgnn}, we evaluate all methods on the link sign prediction task. We report AUC, Binary-F1, Micro-F1, and Macro-F1. AUC measures global ranking ability, but it can remain high even when minority-polarity edges are not well classified. Binary-F1 and Micro-F1 reflect overall classification performance and can be dominated by the majority polarity in highly imbalanced signed graphs. Macro-F1 computes the F1 score for each class separately and takes their unweighted average, making it more informative for class-balanced signed link prediction. For all methods, the positive/negative decision threshold used for the F1 metrics is selected on the validation set by Macro-F1.

\textbf{Training Configuration}. All experiments run on a single RTX 4090D GPU. For Bitcoin-Alpha and Bitcoin-OTC, the batch sizes are 2,048 and 4,096, with learning rates $2\times10^{-4}$ and $5\times10^{-4}$ and an L2 coefficient of $1\times10^{-3}$. For Wiki-RfA, Slashdot, and Epinions, the batch sizes are 8,192, 16,384, and 16,384, with learning rates $\{1,5,5\}\times10^{-4}$ and L2 coefficients $\{5,5,1\}\times10^{-3}$. For all datasets, we set $\delta=1\times10^{-4}$, $\gamma=0.2$, the truncation bound $\epsilon=0.05$, the feature dimension and SVD rank to $128$, the number of attention heads to $4$, dropout to $0.2$, the number of signed clusters to $2$, and a random edge-masking rate of $0.2$. We train with Adam for up to $500$ epochs and stop early on validation AUC with a patience of $30$, and average all reported results over five random seeds. On the above hardware, the average training time per run ranges from about 35 seconds on Bitcoin-OTC to roughly 20 minutes on Epinions.

To avoid information leakage, all structure-dependent preprocessing, including truncated SVD, signed clustering, structural-gradient computation, and prior-score estimation, is fitted only on the training graph. Validation and test edges are held out from these steps and used only for model selection and final evaluation.

\subsection{Overall Performance Comparison}

\begin{table*}[t]
  \caption{Performance comparison on five real-world datasets (mean $\pm$ standard deviation). The PASC column reports the variant chosen by the pre-defined training-graph rule used in Fig.~\ref{fig:3}.}
  \label{tab:performance}
  \centering
  \setlength{\tabcolsep}{3pt}
  \footnotesize 
  \begin{tabular}{llccccccccc}
    \toprule
    Dataset & Metrics & nSNE & SGCN & SNEA & SDGNN & RSGNN & SPMF & TrustSGCN & CFSE & PASC \\
    \midrule
    \multirow{4}{*}{\centering Bitcoin-Alpha} 
    & AUC & 84.25$\pm$2.72 & 88.57$\pm$0.19 & 89.20$\pm$1.60 & 90.29$\pm$1.27 & 79.68$\pm$0.71 & 90.15$\pm$0.28 & 86.72$\pm$0.93 & 93.37$\pm$0.40 & \textbf{93.61$\pm$0.53} \\
    & Binary-F1 & 94.07$\pm$5.01 & 97.07$\pm$0.05 & 93.60$\pm$0.47 & 97.23$\pm$0.30 & 94.81$\pm$0.08 & 97.15$\pm$0.11 & 97.00$\pm$0.04 & 97.81$\pm$0.02 & \textbf{97.85$\pm$0.03} \\
    & Micro-F1 & 89.37$\pm$8.30 & 94.35$\pm$0.11 & 88.55$\pm$0.81 & 94.75$\pm$0.58 & 90.56$\pm$0.15 & 94.61$\pm$0.21 & 94.30$\pm$0.06 & 95.84$\pm$0.04 & \textbf{95.95$\pm$0.06} \\
    & Macro-F1 & 57.76$\pm$9.71 & 60.28$\pm$0.92 & 69.44$\pm$1.61 & 72.91$\pm$4.32 & 71.18$\pm$0.49 & 73.24$\pm$0.11 & 69.00$\pm$0.93 & 79.75$\pm$0.32 & \textbf{81.30$\pm$0.36} \\
    \midrule
    \multirow{4}{*}{\centering Bitcoin-OTC} 
    & AUC & 90.02$\pm$0.97 & 91.63$\pm$0.09 & 87.90$\pm$0.68 & 89.69$\pm$1.12 & 81.04$\pm$0.30 & 89.33$\pm$0.40 & 88.92$\pm$0.54 & \textbf{95.00$\pm$0.81} & 94.98$\pm$0.04 \\
    & Binary-F1 & 96.25$\pm$0.42 & 96.21$\pm$0.13 & 92.31$\pm$0.57 & 95.97$\pm$0.33 & 92.69$\pm$0.27 & 95.77$\pm$0.03 & 95.57$\pm$0.28 & 97.34$\pm$0.11 & \textbf{97.53$\pm$0.13} \\
    & Micro-F1 & 93.02$\pm$0.83 & 93.07$\pm$0.21 & 86.80$\pm$0.93 & 92.70$\pm$0.59 & 87.38$\pm$0.42 & 92.31$\pm$0.04 & 91.92$\pm$0.54 & 95.16$\pm$0.22 & \textbf{95.51$\pm$0.23} \\
    & Macro-F1 & 73.44$\pm$5.85 & 78.01$\pm$0.18 & 72.85$\pm$1.35 & 78.48$\pm$1.92 & 73.22$\pm$0.35 & 76.41$\pm$0.11 & 75.14$\pm$2.18 & 85.09$\pm$0.71 & \textbf{86.68$\pm$0.55} \\
    \midrule
    \multirow{4}{*}{\centering Wiki-RfA} 
    & AUC & 87.32$\pm$0.35 & 84.45$\pm$0.32 & 86.87$\pm$0.17 & 84.66$\pm$1.82 & 68.11$\pm$0.78 & 84.33$\pm$0.03 & 84.57$\pm$0.09 & 88.21$\pm$0.23 & \textbf{89.28$\pm$0.04} \\
    & Binary-F1 & 90.21$\pm$0.40 & 88.78$\pm$0.55 & 86.89$\pm$0.12 & 89.45$\pm$0.52 & 87.81$\pm$1.12 & 89.45$\pm$0.01 & 88.97$\pm$0.05 & 90.02$\pm$0.33 & \textbf{90.03$\pm$0.08} \\
    & Micro-F1 & 83.95$\pm$0.85 & 80.71$\pm$1.22 & 80.72$\pm$0.18 & 82.91$\pm$0.95 & 80.52$\pm$1.37 & 82.78$\pm$0.01 & 82.14$\pm$0.08 & 84.16$\pm$0.43 & \textbf{84.63$\pm$0.11} \\
    & Macro-F1 & 72.81$\pm$2.58 & 59.93$\pm$5.64 & 75.23$\pm$0.24 & 72.27$\pm$2.12 & 69.56$\pm$0.15 & 71.33$\pm$0.04 & 71.05$\pm$0.08 & 75.79$\pm$0.23 & \textbf{78.33$\pm$0.11} \\
    \midrule
    \multirow{4}{*}{\centering Slashdot} 
    & AUC & 88.26$\pm$1.19 & 88.43$\pm$0.16 & 88.98$\pm$0.35 & 87.22$\pm$0.60 & 78.35$\pm$0.12 & 89.13$\pm$0.03 & 88.01$\pm$0.21 & \textbf{93.28$\pm$0.96} & 93.22$\pm$0.03 \\
    & Binary-F1 & 90.63$\pm$0.72 & 90.27$\pm$0.10 & 87.51$\pm$0.20 & 89.83$\pm$0.20 & 87.17$\pm$0.76 & 90.88$\pm$0.02 & 89.94$\pm$0.11 & \textbf{92.51$\pm$0.63} & 92.48$\pm$0.08 \\
    & Micro-F1 & 84.62$\pm$1.32 & 83.84$\pm$0.23 & 81.91$\pm$0.26 & 84.13$\pm$0.34 & 81.14$\pm$0.89 & 85.62$\pm$0.02 & 84.14$\pm$0.16 & 88.35$\pm$0.91 & \textbf{88.48$\pm$0.08} \\
    & Macro-F1 & 73.86$\pm$3.08 & 71.34$\pm$0.85 & 77.33$\pm$0.27 & 76.86$\pm$0.59 & 75.77$\pm$0.61 & 78.45$\pm$0.03 & 76.28$\pm$0.23 & 83.17$\pm$1.08 & \textbf{83.94$\pm$0.04} \\
    \midrule
    \multirow{4}{*}{\centering Epinions} 
    & AUC & 94.59$\pm$0.38 & 93.34$\pm$2.42 & 93.21$\pm$0.16 & 94.27$\pm$0.21 & 87.92$\pm$0.11 & 95.23$\pm$0.03 & 92.91$\pm$0.09 & 97.81$\pm$0.35 & \textbf{97.88$\pm$0.02} \\
    & Binary-F1 & 96.23$\pm$0.22 & 95.65$\pm$1.07 & 93.59$\pm$0.06 & 96.41$\pm$0.35 & 94.84$\pm$0.06 & 96.74$\pm$0.05 & 95.84$\pm$0.06 & 97.62$\pm$0.21 & \textbf{97.71$\pm$0.03} \\
    & Micro-F1 & 93.42$\pm$0.38 & 92.28$\pm$2.05 & 89.41$\pm$0.10 & 93.80$\pm$0.62 & 91.38$\pm$0.10 & 94.38$\pm$0.02 & 92.79$\pm$0.10 & 95.90$\pm$0.35 & \textbf{96.06$\pm$0.04} \\
    & Macro-F1 & 85.14$\pm$0.87 & 80.47$\pm$7.47 & 81.59$\pm$0.09 & 86.69$\pm$1.42 & 84.39$\pm$0.16 & 88.18$\pm$0.05 & 84.47$\pm$0.26 & 91.48$\pm$0.75 & \textbf{91.89$\pm$0.07} \\
    \bottomrule
  \end{tabular}
\end{table*}

For RQ1, Table~\ref{tab:performance} compares PASC with eight baselines on the LSP task. PASC achieves the best Macro-F1 on all datasets while remaining competitive on AUC, Binary-F1, and Micro-F1. This matches the goal of PASC, whose main benefit is not global ranking but class-balanced prediction under polarity asymmetry, where errors on minority-polarity and structurally conflicting edges are more visible in Macro-F1.

The gains are clearest on the most imbalanced datasets. On Bitcoin-Alpha and Bitcoin-OTC, where positive edges outnumber negative ones by 14.7:1 and 9.0:1, PASC improves Macro-F1 over the second-best baseline by 1.55\% and 1.59\%. On Wiki-RfA the Macro-F1 gain reaches 2.54\%, far above the corresponding Micro-F1 and Binary-F1 gains (0.47\% and 0.01\%), suggesting that the improvement comes mainly from class-balanced prediction rather than majority-dominated accuracy.

Several baselines obtain high Binary-F1 or Micro-F1 but much lower Macro-F1, indicating that their overall performance is partly dominated by the majority polarity. For instance, SGCN obtains 59.93\% Macro-F1 on Wiki-RfA, and nSNE reaches 57.76$\pm$9.71\% Macro-F1 on Bitcoin-Alpha. Even CFSE, the strongest baseline, which slightly outperforms PASC in AUC on Slashdot, obtains lower Macro-F1 and relies on explicit closed-subgraph extraction for each target pair. In contrast, PASC keeps Macro-F1 stable across datasets, with standard deviations at most 0.55\%, supporting target-edge structural-prior calibration for class-balanced signed link prediction.

\subsection{Variant Adaptability Analysis}

\begin{figure}[htbp]
  \centering
  \includegraphics[width=1\linewidth]{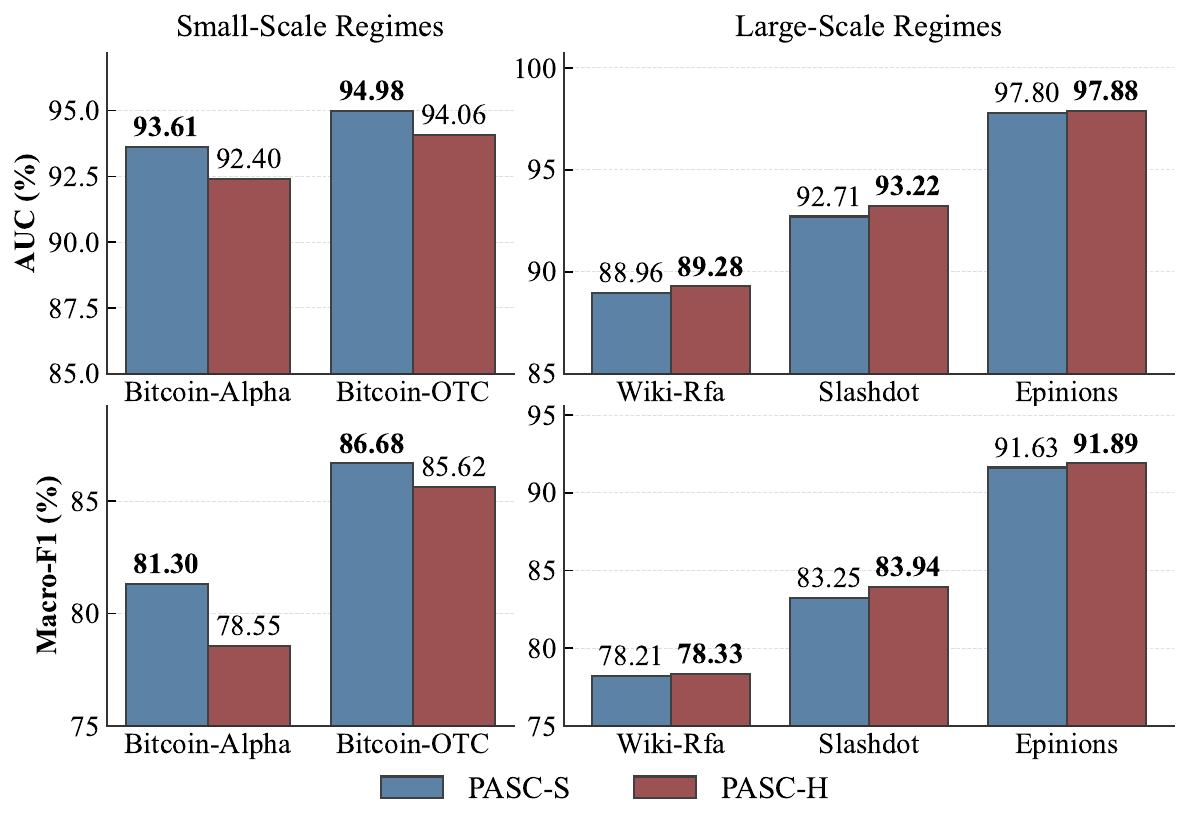}
  \caption{Variant Adaptability comparison of PASC-S and PASC-H under different signed-topology regimes.}
  \label{fig:3}
\end{figure}

For RQ2, Figure~\ref{fig:3} compares the two residual-optimization variants. PASC-S is best on the two small-scale Bitcoin networks on both AUC and Macro-F1, while PASC-H is best on Wiki-RfA, Slashdot, and Epinions. This matches the two designs. PASC-S applies soft residual-aware reweighting and is more stable when local evidence and matched candidate pairs are limited, whereas PASC-H relies on residual-contrastive twin supervision, which needs enough structurally similar but context-opposite pairs and therefore benefits from larger graphs.

The comparison also shows that PASC-H is not uniformly stronger. On small-scale graphs its stronger supervision can disturb limited evidence, giving 2.75\% lower Macro-F1 than PASC-S on Bitcoin-Alpha, while on large-scale graphs the two variants stay within 0.69\% Macro-F1. We therefore pre-select the variant from the training graph size alone, using PASC-S when $|\mathcal{E}_{train}|<5\times10^4$ and PASC-H otherwise. No dataset lies near this threshold, so the rule is insensitive to its exact value and uses no validation or test information.

\subsection{Ablation Study}

\begin{figure}[htbp]
  \centering
  \includegraphics[width=1\linewidth]{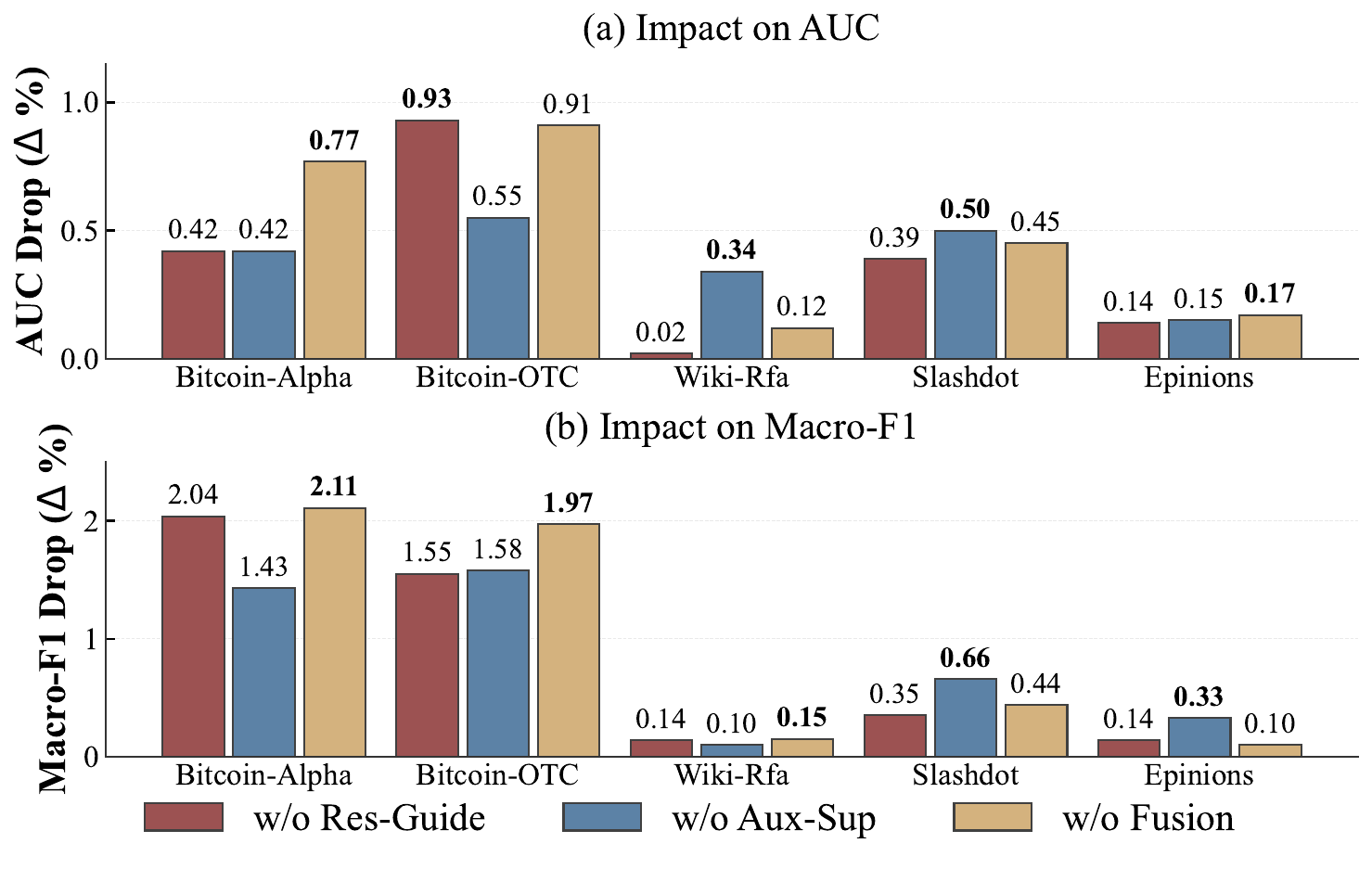}
    \caption{Ablation study of PASC. Panels (a) and (b) report the degradation ($\Delta\%$) in AUC and Macro-F1 when each component is removed.}
  \label{fig:4}
\end{figure}

For RQ3, we evaluate three ablation variants and report their AUC and Macro-F1 degradation.
\begin{itemize}
    \item \textbf{w/o Res-Guide} removes the residual-guided attention aggregation mechanism
    \item \textbf{w/o Aux-Sup} reduces each variant's optimization loss to a plain BCE objective, dropping the residual-aware reweighting on the PASC-S datasets (Bitcoin-Alpha, Bitcoin-OTC) and the auxiliary supervision on the PASC-H datasets (Wiki-RfA, Slashdot, Epinions)
    \item \textbf{w/o Fusion} replaces gated fusion with concatenation
\end{itemize}

Figure~\ref{fig:4} shows that removing each component causes non-zero degradation on all datasets, but its effect depends on the graph regime. On the small-scale PASC-S datasets, removing any component causes a clear drop, up to 2.11\% in Macro-F1 on Bitcoin-Alpha while AUC falls by at most 0.77\%. Since AUC can stay high under class imbalance, this gap links the components mainly to minority and structurally difficult edges rather than overall ranking. The large-scale PASC-H datasets have richer and more redundant structure, so the full model is robust and individual components matter less, with Macro-F1 drops staying below 0.7\%. Even so, the auxiliary supervision remains the most influential component, giving the largest Macro-F1 drop on Slashdot (0.66\%) and Epinions (0.33\%). The three components therefore improve class-balanced prediction, matter most when local evidence is scarce, and leave residual-driven optimization as the dominant component in large-scale regimes.

\subsection{Hyperparameter Sensitivity Analysis}

\begin{figure}[htbp]
  \centering
  \includegraphics[width=1.0\linewidth]{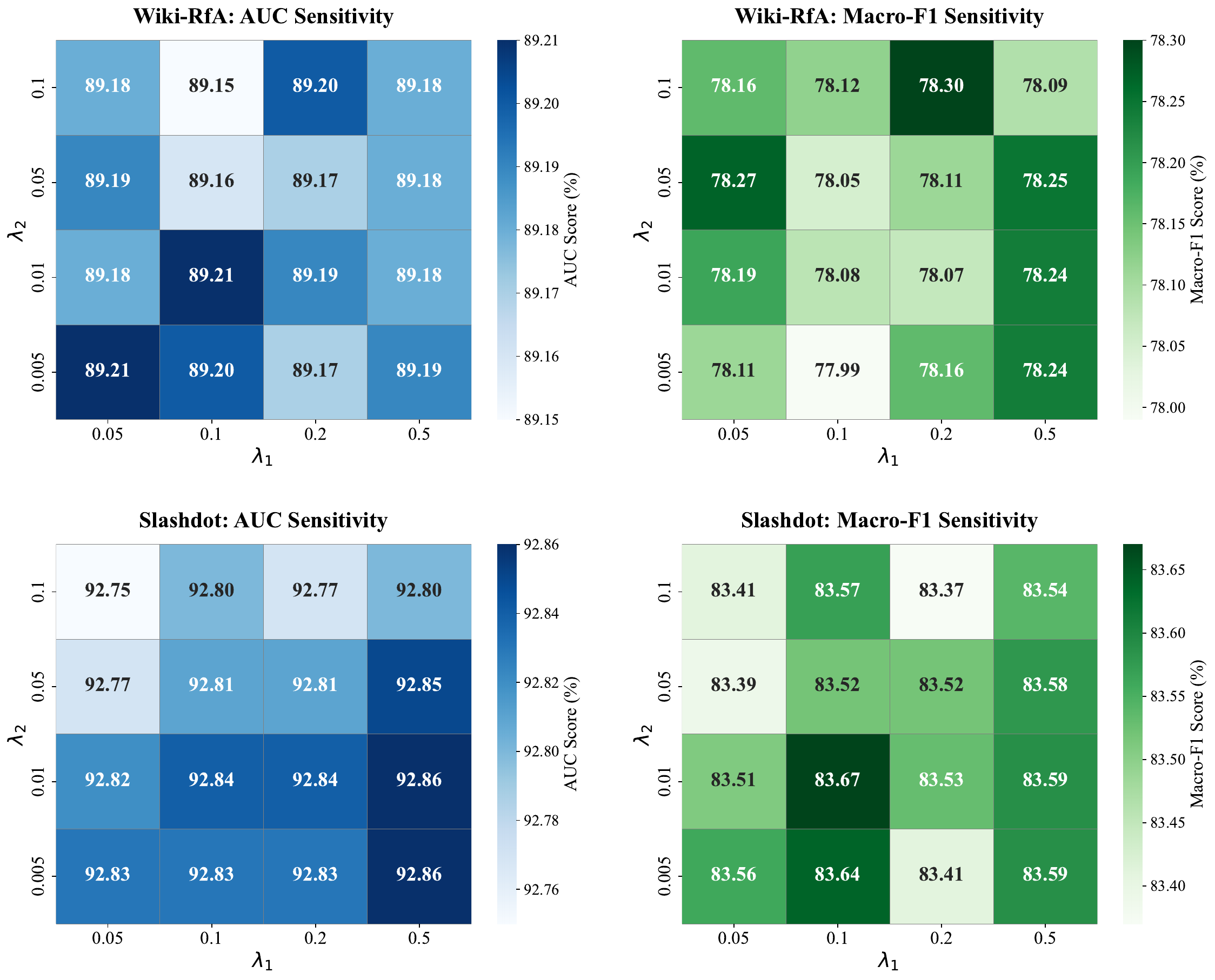}
  \caption{Sensitivity analysis of $\lambda_1$ and $\lambda_2$ on Wiki-RfA and Slashdot, showing stable AUC and Macro-F1 across loss-weight settings.}
  \label{fig:5}
\end{figure}

For RQ4, we perform a grid search on two representative PASC-H datasets, Wiki-RfA and Slashdot, with $\lambda_1 \in \{0.05, 0.1, 0.2, 0.5\}$ and $\lambda_2 \in \{0.005, 0.01, 0.05, 0.1\}$. As shown in Figure~\ref{fig:5}, PASC-H remains stable across the tested loss-weight settings, with AUC varying by at most 0.11\% and Macro-F1 by at most 0.31\% over the entire grid. The gain of PASC-H therefore does not depend on a finely tuned loss-weight combination, but holds across a reasonable range of auxiliary-loss weights.

\begin{table*}[!t]
  \caption{Performance comparison under the Degree-Shift structural-shift protocol. For space, structural-shift tables report AUC and Macro-F1. Binary-F1 and Micro-F1 show similar trends and are omitted.}
  \label{tab:Degree}
  \centering
  \footnotesize 
  \setlength{\tabcolsep}{3pt}
  \begin{tabular}{llccccccccc}
    \toprule
    Dataset & Metrics & nSNE & SGCN & SNEA & SDGNN & RSGNN & SPMF & TrustSGCN & CFSE & PASC \\
    \midrule
    \multirow{2}{*}{\centering Bitcoin-Alpha} 
    & AUC       & 69.53$\pm$2.93 & 80.60$\pm$1.58 & 83.10$\pm$0.70 & 80.71$\pm$0.81 & 74.31$\pm$0.04 & 83.40$\pm$0.54 & 79.89$\pm$0.23 & 77.94$\pm$1.39 & \textbf{84.52$\pm$0.17} \\
    & Macro-F1  & 49.80$\pm$1.97 & 54.89$\pm$6.16 & 70.17$\pm$0.58 & 63.19$\pm$3.48 & 66.61$\pm$0.83 & 71.16$\pm$0.19 & 59.46$\pm$1.93 & 60.92$\pm$2.22 & \textbf{72.05$\pm$0.35} \\
    \midrule
    \multirow{2}{*}{\centering Bitcoin-OTC} 
    & AUC       & 76.48$\pm$7.61 & 88.01$\pm$0.10 & 85.46$\pm$0.10 & 87.01$\pm$0.13 & 79.34$\pm$0.18 & 88.10$\pm$0.35 & 87.77$\pm$0.09 & 78.26$\pm$5.80 & \textbf{89.09$\pm$0.03} \\
    & Macro-F1  & 53.46$\pm$7.69 & 75.71$\pm$0.14 & 74.66$\pm$0.28 & 75.67$\pm$0.11 & 71.61$\pm$0.28 & 76.75$\pm$0.25 & 74.26$\pm$0.20 & 68.10$\pm$3.47 & \textbf{80.06$\pm$0.26} \\
    \midrule
    \multirow{2}{*}{\centering Wiki-RfA} 
    & AUC       & 69.46$\pm$1.30 & 78.81$\pm$1.19 & 79.36$\pm$0.16 & 77.27$\pm$0.87 & 66.62$\pm$1.17 & 78.84$\pm$0.15 & 77.08$\pm$3.01 & 72.25$\pm$4.89 & \textbf{79.70$\pm$0.13} \\
    & Macro-F1  & 50.08$\pm$9.25 & 67.02$\pm$1.25 & 66.13$\pm$0.05 & 69.23$\pm$0.21 & 53.81$\pm$0.62 & 69.23$\pm$0.18 & 65.48$\pm$3.83 & 62.83$\pm$1.48 & \textbf{69.24$\pm$0.21} \\
    \midrule
    \multirow{2}{*}{\centering Slashdot} 
    & AUC       & 72.27$\pm$3.78 & 83.29$\pm$0.45 & 80.23$\pm$1.07 & 78.63$\pm$0.37 & 73.71$\pm$0.81 & 79.62$\pm$0.16 & 82.11$\pm$1.19 & \textbf{85.30$\pm$0.33} & 85.15$\pm$0.04 \\
    & Macro-F1  & 56.69$\pm$6.64 & 70.97$\pm$2.60 & 71.09$\pm$0.73 & 66.71$\pm$0.96 & 71.32$\pm$0.25 & 69.56$\pm$0.23 & 67.05$\pm$0.98 & 74.10$\pm$0.66 & \textbf{74.81$\pm$0.12} \\
    \midrule
    \multirow{2}{*}{\centering Epinions} 
    & AUC       & 84.28$\pm$1.79 & 89.90$\pm$0.39 & 86.35$\pm$0.18 & 90.05$\pm$0.06 & 83.12$\pm$0.21 & 89.65$\pm$0.13 & 88.18$\pm$0.14 & 88.81$\pm$1.83 & \textbf{92.42$\pm$0.27} \\
    & Macro-F1  & 72.30$\pm$6.78 & 76.80$\pm$3.61 & 73.26$\pm$0.85 & 81.28$\pm$0.02 & 76.19$\pm$0.24 & 80.12$\pm$0.04 & 74.77$\pm$0.72 & 81.61$\pm$0.79 & \textbf{82.38$\pm$0.72} \\
    \bottomrule
  \end{tabular}
\end{table*}

\begin{table*}[t]
  \caption{Performance comparison under the Structural-Shortcut-Shift structural-shift protocol. For space, structural-shift tables report AUC and Macro-F1. Binary-F1 and Micro-F1 show similar trends and are omitted.}
  \label{tab:Structural}
  \centering
  \footnotesize 
  \setlength{\tabcolsep}{3pt}
  \begin{tabular}{llccccccccc}
    \toprule
    Dataset & Metrics & nSNE & SGCN & SNEA & SDGNN & RSGNN  & SPMF & TrustSGCN & CFSE & PASC \\
    \midrule
    \multirow{2}{*}{\centering Bitcoin-Alpha} 
    & AUC       & 66.02$\pm$5.45 & 79.98$\pm$0.70 & 79.17$\pm$0.87 & 79.77$\pm$1.50 & 73.37$\pm$0.35 & 79.00$\pm$0.43 & 80.09$\pm$2.12 & 76.10$\pm$1.27 & \textbf{81.67$\pm$0.15} \\
    & Macro-F1  & 49.44$\pm$1.68 & 60.74$\pm$1.07 & 65.82$\pm$0.62 & 62.49$\pm$3.51 & 62.19$\pm$0.11 & 65.48$\pm$0.49 & 60.38$\pm$2.60 & 59.48$\pm$1.97 & \textbf{67.53$\pm$0.38} \\
    \midrule
    \multirow{2}{*}{\centering Bitcoin-OTC} 
    & AUC       & 73.00$\pm$2.18 & 83.86$\pm$0.21 & 82.79$\pm$2.38 & 83.87$\pm$0.21 & 76.38$\pm$0.02 & 83.20$\pm$0.95 & 84.05$\pm$0.21 & 80.66$\pm$1.99 & \textbf{85.33$\pm$0.10} \\
    & Macro-F1  & 59.85$\pm$5.63 & 70.81$\pm$0.71 & 72.80$\pm$1.57 & 69.35$\pm$0.45 & 69.68$\pm$0.21 & 71.84$\pm$0.47 & 65.48$\pm$0.93 & 69.90$\pm$0.92 & \textbf{74.36$\pm$0.54} \\
    \midrule
    \multirow{2}{*}{\centering Wiki-RfA} 
    & AUC       & 80.28$\pm$2.19 & 80.52$\pm$0.10 & 84.12$\pm$0.20 & 81.99$\pm$0.62 & 71.48$\pm$0.10 & 80.95$\pm$0.07 & 80.47$\pm$0.71 & 83.00$\pm$0.87 & \textbf{84.19$\pm$0.23} \\
    & Macro-F1  & 44.53$\pm$4.91 & 69.06$\pm$2.20 & 75.08$\pm$0.28 & 71.66$\pm$0.68 & 71.91$\pm$0.32 & 73.66$\pm$0.07 & 64.89$\pm$0.39 & 73.95$\pm$0.82 & \textbf{75.28$\pm$0.51} \\
    \midrule
    \multirow{2}{*}{\centering Slashdot} 
    & AUC       & 70.43$\pm$1.71 & 82.28$\pm$0.12 & 77.56$\pm$0.33 & 78.34$\pm$0.06 & 73.77$\pm$0.28 & 78.97$\pm$0.20 & 82.36$\pm$0.83 & 83.20$\pm$1.00 & \textbf{84.04$\pm$0.11} \\
    & Macro-F1  & 61.09$\pm$0.41 & 65.96$\pm$3.41 & 69.28$\pm$0.26 & 66.59$\pm$0.12 & 70.98$\pm$0.39 & 69.00$\pm$0.03 & 68.20$\pm$1.46 & 73.42$\pm$2.19 & \textbf{74.07$\pm$0.17} \\
    \midrule
    \multirow{2}{*}{\centering Epinions} 
    & AUC       & 83.53$\pm$0.73 & 87.59$\pm$0.07 & 85.05$\pm$0.38 & 89.02$\pm$0.16 & 80.13$\pm$0.42 & 88.69$\pm$0.12 & 87.94$\pm$0.16 & 89.89$\pm$0.74 & \textbf{91.93$\pm$0.12} \\
    & Macro-F1  & 68.28$\pm$11.75 & 61.18$\pm$2.02 & 77.40$\pm$0.48 & 80.59$\pm$0.31 & 79.14$\pm$0.23 & 80.75$\pm$0.02 & 73.97$\pm$0.01 & 82.51$\pm$0.38 & \textbf{84.42$\pm$0.08} \\
    \bottomrule
  \end{tabular}
\end{table*}

\subsection{Structural-Shift Robustness Analysis}
For RQ5, we evaluate PASC under two structural-shift protocols, Degree-Shift and Structural-Shortcut-Shift, to test whether PASC remains robust when the structural properties of test edges differ from those of training edges. Degree-Shift tests robustness from dense, high-degree regions to sparse, low-degree regions. Structural-Shortcut-Shift tests whether models remain effective when test edges contain weaker local closure patterns, reducing the availability of common-neighbor-based structural shortcuts.

\textbf{Degree-Shift}. We define the edge density score as:
\begin{equation}
    s_{\text{deg}}(u,v)=d(u)\cdot d(v),
\end{equation}
where $d(\cdot)$ denotes node degree. After sorting $s_{\text{deg}}$ in descending order, we use the top 40\%, middle 10\%, and bottom 50\% as the training, validation, and test sets, respectively.

\textbf{Structural-Shortcut-Shift}. We define the local closure score as:
\begin{equation}
    s_{\text{cn}}(u,v)=|N(u)\cap N(v)|,
\end{equation}
where $s_{\text{cn}}(\cdot,\cdot)$ counts the number of common neighbors. Edges with $s_{\text{cn}}\ge3$, $s_{\text{cn}}=2$, and $s_{\text{cn}}\le1$ are used for training, validation, and testing, respectively. If the middle bucket is too small for validation, we sample additional validation edges from the training domain without using any test-domain edges.

Under Degree-Shift, Table~\ref{tab:Degree} shows that most baselines degrade substantially when test edges move to lower-degree regions with fewer aggregation signals. PASC achieves the best Macro-F1 on all five datasets and the best or comparable AUC. For example, on Bitcoin-Alpha nSNE drops to 49.80\% Macro-F1 while PASC keeps 72.05\%, and on Epinions PASC reaches 82.38\% Macro-F1, 0.77\% above the second-best CFSE. PASC therefore shows reduced dependence on dense local neighborhoods under the tested Degree-Shift protocol.

Structural-Shortcut-Shift more directly tests reliance on local-closure shortcuts, and here PASC shows a clearer advantage. As shown in Table~\ref{tab:Structural}, when test edges have weaker common-neighbor closure, PASC obtains the best AUC and Macro-F1 on all five datasets (for example, 84.42\% Macro-F1 on Epinions versus 82.51\% for CFSE). This supports the claim that calibrating target-edge structural reliance helps when local-closure shortcuts become less available. These protocols change the structural properties of test edges and may also shift the polarity distribution, so we use Macro-F1 as the primary metric to limit majority-polarity dominance.

\section{Conclusion}

We proposed PASC, a target-edge structural-prior calibration framework for link sign prediction in polarity-asymmetric signed networks. PASC estimates structural prior scores, compares them with local signed-context cues, and uses the resulting conflict residuals to guide signed attention aggregation, gated fusion, and regime-adaptive optimization. By calibrating the reliance on signed-graph structural priors for each target edge, PASC improves robustness to unreliable priors while preserving useful local context. Experiments on five real-world signed networks show consistent Macro-F1 gains with competitive overall performance. Structural-shift experiments further suggest reduced dependence on dense-neighborhood and local-closure shortcuts under the tested protocols. Future work will explore more efficient residual estimation and extend PASC to heterogeneous and dynamic signed networks.

\section*{Acknowledgments}
This work was supported by the National Natural Science Foundation of China under Grant 62307006 and 62676166.

\bibliographystyle{IEEEtran}
\bibliography{ref}

\end{document}